Jun Chen,[1,2,3]* Jack Ng,[1,4]*† Kun Ding,[1,5] Kin Hung Fung,[6]

Zhifang Lin,[3] and C. T. Chan[1,5]‡

[1] *Department of Physics, The Hong Kong University of Science and Technology*

[2] *Institute of Theoretical Physics and Department of Physics, Shanxi University*

[3] *State Key Laboratory of Surface Physics, Key Laboratory of Micro and Nano Photonic Structures (MOE), and Department of physics, Fudan University*

[4] *Department of Physics and Institute of Computational and Theoretical Studies, Hong Kong Baptist University*

[5] *Institute for Advanced Study, The Hong Kong University of Science and Technology*

[6] *Department of Applied Physics, Hong Kong Polytechnic University*

* These authors contributed equally.

† jacktfng@hkbu.edu.hk

‡ phchan@ust.hk

**Negative Optical Torque**

Maxwell noted that light carries angular momentum[1,2], and as such it can exert torques on material objects[3]. This was subsequently proved by Beth in 1936[4]. Applications of these opto-mechanical effects were limited initially due to their smallness in magnitude, but later enabled by the invention of laser. Novel[5-12] and practical[13-20] approaches for harvesting light for particle rotation have been demonstrated, where the structure is subjected to a positive optical torque along a certain axis[21] if the incident angular momentum has a positive projection on the same axis. We report here a counter-intuitive phenomenon of "negative optical torque", meaning that incoming photons carrying angular momentum rotate an object in the opposite sense. Surprisingly this can be realized quite straightforwardly in simple planar structures. Field retardation is a necessary condition. The optimal conditions are explored and explained.

The negative optical torque (NOT) phenomenon manifestly conserves total angular momentum (AM). Let us consider a structure illuminated by a wave carrying a $z$-component AM of $L_z^{inc}$. After being scattered by the structure, the total $z$-component AM of the light is $L_z^{sca}$. If the consequence of scattering is such that $L_z^{sca} > L_z^{inc}$, then AM conservation requires that their difference must be balanced by a NOT acting on the structure. Consequently, albeit somewhat counter-intuitive, NOT does not violate AM conservation. The question is then whether such phenomenon, which does not violate conservation laws, can be realized in practice. If so, is NOT easily observable? We will show that NOT is in fact ubiquitous, as it can be realized in many transparent or weakly absorptive structures, in discrete rotationally symmetric or even irregular structures, and also in planar or even 3D structures (see session 4 of supporting information). We shall show that field retardation is necessary for achieving NOT and discrete rotational symmetry (DRS) enhances NOT. The realization of NOT may open up new applications in areas where optical micromanipulation play a role[22-23]. We remark that NOT is the angular counterpart of the phenomenon of optical pulling force[24-27].

Explicit examples of structure where NOT can be realized are shown in Fig. 1. For the ease and accuracy of computation and interpretation, we consider planar structures composing of dielectric microspheres located on the *xy*-plane. The axis for which the torque is to be calculated is parallel to the *z*-axis and goes through the center of mass marked in Fig.1, where some of the structures possess DRS but some do not. We stress that our conclusions are valid not only for discrete system comprising of spheres, but also to a continuous piece of material.

The time-averaged optical torque $\boldsymbol{\Gamma}$ acting on the concerned structure is computed by (see supplementary information)

$$\boldsymbol{\Gamma} = \iint_{\sigma} \left( \mathbf{r} \times \ddot{\mathbf{T}} \right) \cdot \hat{\mathbf{n}} dS . \qquad (1)$$

where the origin of the coordinate system is chosen to be the center of mass marked in Fig. 1. $\ddot{\mathbf{T}}$ is the time averaged Maxwell stress tensor (see supplementary information), $\sigma$ is any surface that encloses the entire structure, and $\mathbf{r}$ is the position vector originates from the center of mass. To evaluate $\ddot{\mathbf{T}}$, the generalized Mie scattering theory for multi-sphere[28] is employed. Irrespective of the structural orientation and morphology, the *z*-component optical torque is given by

$$\Gamma_z = \left( \sum_{m_i} m_i \tau_{m_i}^{(\text{Extinct})} - \sum_{m} m \tau_{m}^{(\text{Recoil})} \right), \qquad (2)$$

where $\omega \tau_{m}^{(\text{Extinct})}$ and $\omega \tau_{m}^{(\text{Recoil})}$ are, respectively, the energy extinction rate by the particle via channel *m* and the rate in which energy is scattered into the channel *m* (both are tabulated in the *Methodology* session), $\omega$ is the angular frequency of incident wave, and $m_i$ and *m* characterize, respectively, the azimuthal channels for the incident and scattered waves. Our approach can be considered as exact within classical electrodynamics (subject only to numerical truncation error), which is highly precise when applied to micro-particles. Unless stated otherwise, the incident wave is a left circularly polarized plane wave propagating along the *z*-axis, with a positive AM and an intensity of $1 \text{ mW}/\mu\text{m}^2$.

Figure 2 shows the optical torques versus *ka* for three morphologically different planar structures with different types of incident waves, where *k* is the wavenumber and *a* is the radius of the comprising spheres. The results clearly show that NOT can be realized in all three structures, which include planar triangular or hexagonal structures consisting of silicon or polystyrene microspheres, illuminated by a circularly polarized plane wave or Gaussian beam. The optical torques can take negative values at some intervals of *ka*. The three structures in Fig. 2 are not of special design. Other planar structures can serve the same purpose. Figure 2 illustrates that NOT is ubiquitous: it can be observed in different structures (planar or non-planar) with different forms of incident wave, including plane wave and Gaussian beam.

We now analytically analyze the physical origin of the seemingly counter-intuitive but in fact ubiquitous NOT. The transfer of AM can be viewed as occurring in two steps. First, the photons are intercepted by the scatterer. This induces an extinction torque given by the first term in Eq. (2). For an incident wave consisting of solely positive $m_i$ components, this extinction torque is positive definite. So if NOT exists at all, it must come from the recoil torque in Eq. (2). Second, some of the intercepted photons will be absorbed, while some will be re-emitted (i.e., scattering). The re-emitted photons exert a recoil toque on the scatterer, as given by the second term in Eq. (2), where *m*<0 (*m*>0) represent a positive (negative) contribution. If the negative contribution dominates, we will have a NOT. A detailed discussion of the extinction and recoil torque is given in the supplementary information.

NOT is closely tied to the symmetry of the structure and incident wave. A rotationally symmetric scatterer (such as a sphere or cylinder) cannot exhibit a NOT along its symmetry axis, because the azimuthal number $m$ is preserved in the scattering process. Then an incident wave with $m_i$ can only be scattered into the same azimuthal channel characterized by the same $m_i$. Eq. (2) reduces to $\Gamma_z = \sum_{m_i} m_i \tau_{m_i}^{(Absrob)}$, where $\tau_{m_i}^{(Absrob)} = \tau_{m_i}^{(Extinct)} - \tau_{m_i}^{(Recoil)}$ is proportional to the absorption in channel $m_i$. Since both $\tau_{m_i}^{(Absrob)}$ and $m_i$ are positive definite, $\Gamma_z \geq 0$. This proves that if NOT is to exist at all, the scatterer must couple different azimuthal channels, which is possible only in non-spherical scatterer. Here, for simplicity, we shall focus on structures with DRS and for an incident wave with a single azimuthal channel $m_i>0$ (but composed of different angular momentum channels [28]). It is known that when the incident wave has both positive and negative $m_i$, if the particle is more responsive to the positive (negative) $m_i$, the torque will be positive (negative). Here, we demonstrate a completely different phenomenon where a positive $m_i$ can induce a NOT.

For scatterers possessing $m_s$-fold DRS, channel $m_i$ is scattered into azimuthal channels characterized by[29] (see session 1 of supporting information):

$$m = m_i + n \times m_s \tag{3}$$

where $n = 0, \pm 1, \pm 2$ and so on. Eq. (3) is the angular analog of the phase matching condition for a periodic surface illuminated by a plane wave. Under these conditions, Eq. (2) can be re-written as

$$\Gamma_z = m_i \tau_{m_i}^{(Extinct)} - \sum_n (m_i + n \times m_s) \tau_{m_i + n \times m_s}^{(Recoil)} \tag{4}$$

For dielectric material, $n$ is not a big number due to the weak multiple scattering (see session 2 of supporting information). In the extreme situation of $m_s \ll m_i$, the torque is positive definite as

$$\Gamma_z \approx m_i \left( \tau_{m_i}^{(\text{Extinct})} - \sum_n \tau_{m_i + n \times m_s}^{(\text{Recoil})} \right) = m_i \left( W_{ext} - W_{sca} \right) / \omega = m_i W_{abs} / \omega \geq 0 \qquad (5)$$

where $W_{ext} = \omega \tau_{m_i}^{(\text{Extinct})}$, $W_{sca} = \omega \sum_n \tau_{m_i + n \times m_s}^{(\text{Recoil})}$, and $W_{abs} = \left( W_{ext} - W_{sca} \right)$ are, respectively, the rates at which energy is intercepted, scattered, and absorbed. When $m_s \gg m_i$, since angular momentum number $l$ is always greater than or equal to $m$[30], $l$ involved in $W_{sca}$ will be very large whenever $n \neq 0$. However waves characterize by a very large $l$ cannot reach the scatterer due to centrifugal effects. It is then clear that if NOT can exist, it must be for some moderate value of $m_s$, where there are partial waves with relatively smaller $l \geq m_s$ that can reach the scatterer to transfer AM. This is in fact a manifestation of the optical diffraction limit. If the small structure has some very fine sub-wavelength details, such as patterns with high degree of rotational symmetry located in a small region, the incident wave will not be able to probe them. Thus the information on the symmetry will be lost, eliminating the possibility of observing NOT.

Next, we argue that retardation is essential for NOT. We show in Fig. 3 the optical torque due to individual azimuthal channel for structures with three-fold ($m_s = 3$) or six-fold ($m_s = 6$) rotational symmetry. According to Eq. (3), for three-fold symmetric structures showed in Fig. 3(a)-(b), light is scattered into azimuthal channels characterized by $m = 3n+1$, where $n = 0, \pm 1, \pm 2, \ldots$ etc. The $m=1$ channel (i.e. $n=0$ in Eq. (3)) does not contribute to torque for lossless sphere. Light scattered into azimuthal channels $m=-2, -5, -8$ and so on will induce positive optical torque, whereas the channels characterized by *m*=4, 7, 10 and so on will induce NOT. As multiple scattering is not strong in dielectrics, the recoil torque is dominated by $n = \pm 1$, corresponding to $m = m_i \pm m_s = -2$ or $4$. The peaks in Fig. 3 for $n = \pm 1$ channels correspond to a Mie type resonance. Due to retardation, the torques induced by the azimuthal channels corresponding to $n = \pm 1$, and also the others, are oscillatory function of the wavelength and structural size. Precisely due to this retardation induced oscillation, the torque can take positive or negative value. The situation for the six-fold symmetric structures showed in Fig. 3(c)-(d) is similar.

In general, NOT is ubiquitous if the structure's $m_s$ fulfill the following criteria: (1) moderate $m_s$ (for examples, $m_s$=5 to $m_i$=1) and (2) $m_s < l_{max}$, where $l_{max}= kR + 4.05(kR)^{1/3} + 2$ is the AM truncation order for the scatterer[29,30], with $R$ being the size of the scatterer measured from the rotation axis.

In summary, NOT, which is the angular analog of the optical pulling force, can be realized without violation of AM conservation. While the optical pulling force is achievable only for special kinds of optical beams acting on some specific particle[24-27], NOT is ubiquitous in simple configurations and can be induced by using a broad class of incident waves. NOT can occur if the recoil torque happens to be negative and larger in magnitude than the extinction torque, which is made possible by retardation effects and enhanced by DRS. We remark that our theory is valid for both spin and orbital AM (i.e. arbitrary $m_i$) of light.

**Methodology**

We shall work in the basis of the vector spherical wave functions (VSWF) $\mathbf{M}_{ml}^{(J)}(k\mathbf{r})$ and $\mathbf{N}_{ml}^{(J)}(k\mathbf{r})$, which are complete orthogonal basis for the free space Maxwell equations in spherical coordinates[30]. They are also the eigenmodes for the operators of the total AM and z-component of AM. Using VSWF, the incident field can be expanded as

$$\mathbf{E}_{inc} = -\sum_{m,l} iE_{ml}\left[p_{ml}\mathbf{N}^{(1)}_{ml} + q_{ml}\mathbf{M}^{(1)}_{ml}\right], \qquad (6)$$

where the expansion coefficients $(p_{ml}, q_{ml})$ can be obtained from overlap integrals between the incident field and VSWF. The scattered field is given by

$$\mathbf{E}_{sca} = \sum_{m,l} iE_{ml}\left[a_{ml}\mathbf{N}^{(3)}_{ml} + b_{ml}\mathbf{M}^{(3)}_{ml}\right]. \qquad (7)$$

where

$$E_{ml} = |E_0|i^l \sqrt{\frac{2l+1}{l(l+1)} \frac{(l-m)!}{(l+m)!}}, \qquad (8)$$

and the expansion coefficients $(a_{ml}, b_{ml})$ are linear function of $(p_{ml}, q_{ml})$ and can only be obtained after solving the scattering problem by the generalized Mie theory for multi-sphere, i.e. matching the standard boundary conditions over the surface of all spheres to obtain a set of linear equations in which one can solve for $(a_{ml}, b_{ml})$.

The recoil torque and extinction torque in the azimuthal channel $m$ are

$$m\tau_m^{(\text{Recoil})} = m\frac{2\pi\varepsilon_0}{k^3}|E_0|^2 \sum_l \left(|a_{ml}|^2 + |b_{ml}|^2\right), \qquad (9)$$

$$m\tau_m^{(\text{Extinct})} = m\frac{2\pi\varepsilon_0}{k^3}|E_0|^2 \sum_l \mathrm{Re}\left(a_{ml}p^*_{ml} + b_{ml}q^*_{ml}\right). \qquad (10)$$

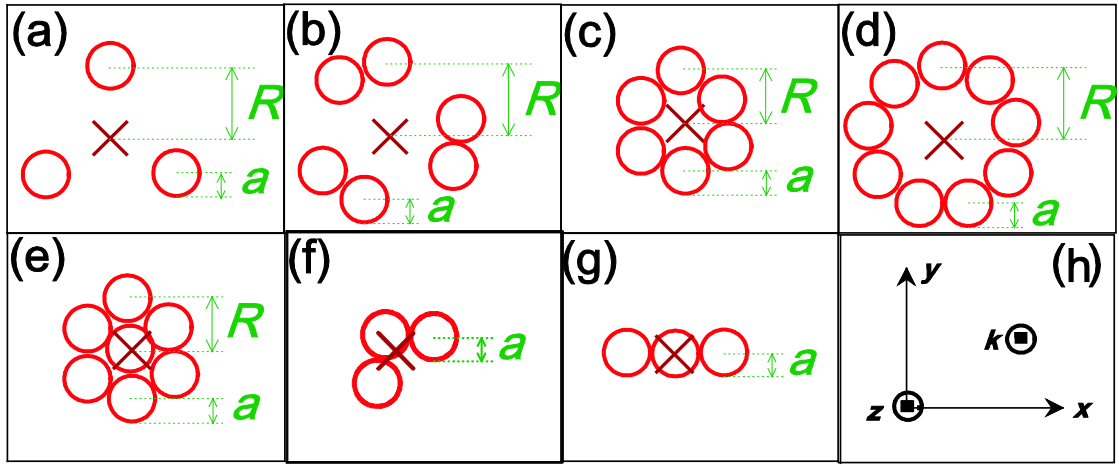

**Figure 1 | Examples of planar structures that can realize NOT.** (a)-(g) structures composed of identical dielectric spheres. $R$ ($a$) is the radius of the structure (sphere). The cross marks the center of mass, which coincides with the axis of rotation. Panel (h) shows the coordinate system. The incident plane wave has a k-vector ($k$) along the z-axis.

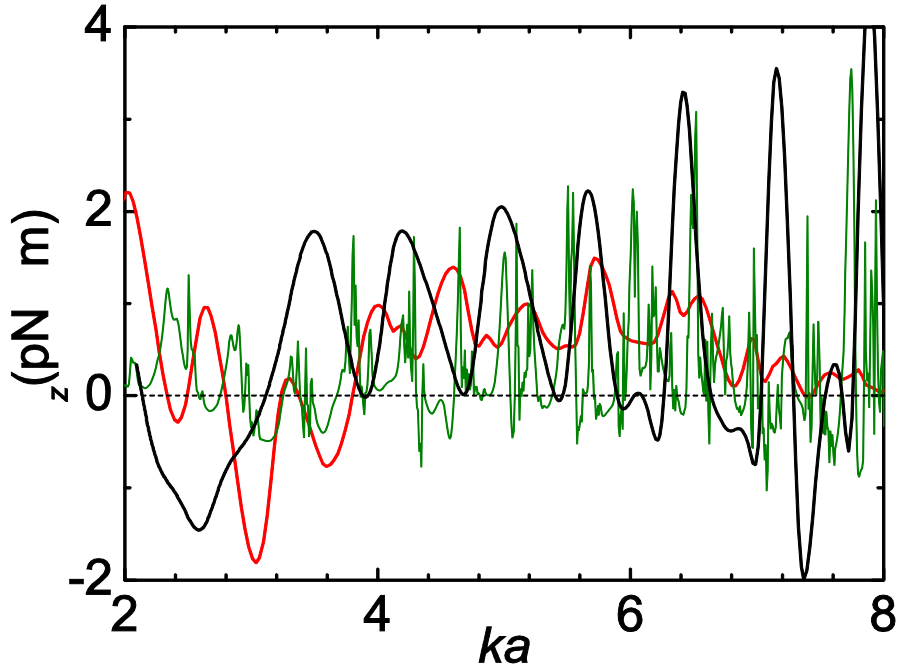

**Figure 2 | Optical torque versus *ka* for morphologically different structures illuminated by different types of incident wave carrying positive angular momentum.** Red: Optical torque for the structure shown in Fig. 1(e) where $k$ is varying ($\varepsilon_r = 2.46$, $a = 0.49\mu m$, $R = 1.0\mu m$). Green: Optical torque for the structure shown in Fig. 1(a) composed of silicon spheres ($\varepsilon_r = 12.6096$, $R = 1.16a$, $\lambda = 1064$nm) in water where $a$ is varying. Black: Optical torque for the structure shown in Fig. 1(a) ($\varepsilon_r = 2.4649$, $a = 0.5\mu m$, $R = 0.58\mu m$) where $k$ is varying. Red and Green curve is associated with a left circularly polarized plane wave of intensity= $1$ mW/$\mu$m$^2$. Black curve is associated with a left circularly polarized Gaussian beam (numerical aperture =0.9 and total power=0.05W). Negative torque ($\tau_z < 0$) means that the light beam induces the object to rotate in the opposite sense as the angular momentum of the incident photon. .

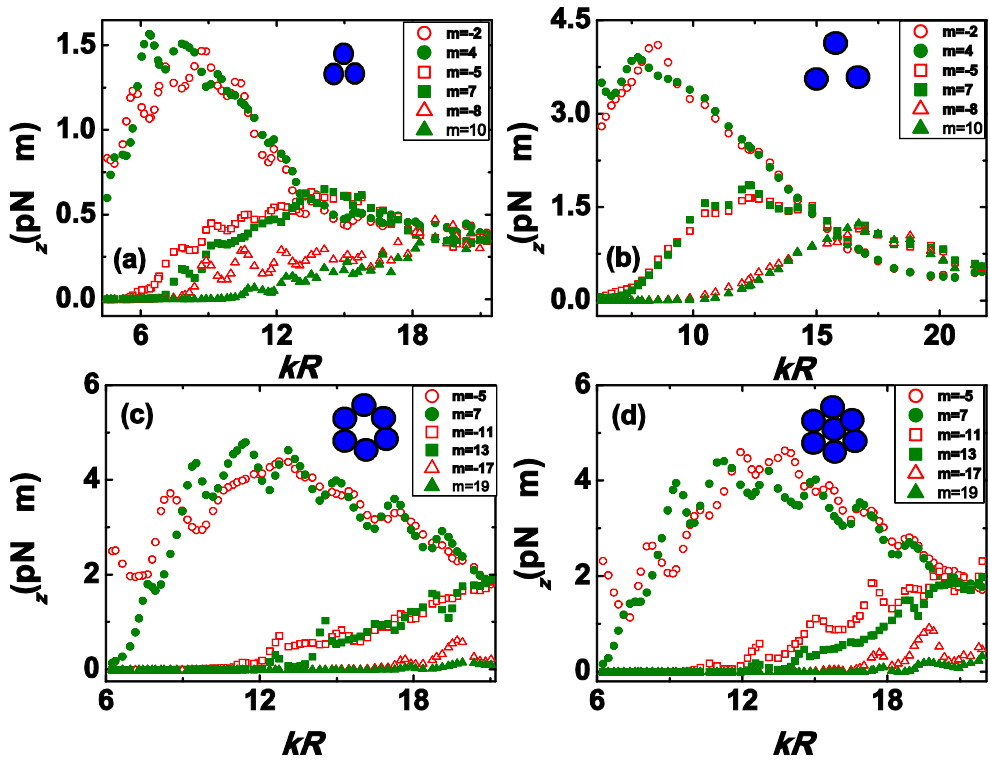

**Figure 3 | The contribution (in absolute value) to the torque by individual partial wave azimuthal channels versus *kR*.** (a)-(b) three-fold and (c)-(d) six-fold rotationally symmetric dielectric structures ($\varepsilon_r = 2.4649$, $a = 0.49 \mu m$). The incident field is a left circularly polarized plane wave and intensity is 1 mW/$\mu$m$^2$. R=0.58μm for (a) and R=1μm for (b)-(d). The oscillations are due to retardation.

**Acknowledgement**

This work is supported by Hong Kong RGC through HKUST2/CRF/11G and AoE/P-02/12. JN is also supported by RGC through HKBU604011 and HKBU603312. ZL was supported by NSFC through 11174059.